# Stabilization of $S_3O_4$ at High Pressure-Implications for the Sulfur Excess Paradox

Siyu Liu[1, †], Pengyue Gao[1, †], Andreas Hermann[2], Guochun Yang[3], Jian Lv[1,*], Yanming Ma[1, 4], Ho-Kwang Mao[5, 6, *], and Yanchao Wang[1, *]

[1] State key laboratory of superhard materials & International center of computational method and software, College of Physics, Jilin University, Changchun 130012, China

[2] Centre for Science at Extreme Conditions and SUPA, School of Physics and Astronomy, The University of Edinburgh, Edinburgh EH9 3FD, United Kingdom

[3] State Key Laboratory of Metastable Materials Science & Technology and Key Laboratory for Microstructural Material Physics of Hebei Province, School of Science, Yanshan University, Qinhuangdao 066004, China

[4] International Center of Future Science, Jilin University, Changchun 130012, China

[5] Center for High Pressure Science and Technology Advanced Research, Beijing 100094, China.

[6] Geophysical Laboratory, Carnegie Institution of Washington, Washington, DC 20015, USA.

**ABSTRACT:** The geological conundrum of "sulfur excess" refers to the finding that predicted amounts of sulfur, in the form of $SO_2$, discharged in volcanic eruptions much exceeds the sulfur available for degassing from the erupted magma. Exploring the source of the excess sulfur has been the subject of considerable interest. Here, from a systematic computational investigation of sulfur-oxygen compounds under pressure, a hitherto unknown $S_3O_4$ compound containing a mixture of sulfur oxidation states +II and +IV emerges and is predicted to be stabilized above a pressure of 79 GPa. We predict that $S_3O_4$ can be produced via multiple redox reactions involving subducted S-bearing minerals (e.g., sulfates and sulfides) at high pressure conditions relevant to the deep lower mantle, and conversely be decomposed into $SO_2$ and S at shallow depths of Earth. Therefore, $S_3O_4$ can be considered as a key intermediate compound to promote the decomposition of sulfates to release $SO_2$, which offers an alternative source of the excess sulfur released during explosive eruptions. These findings provide a possible resolution to the geological paradox of "excess sulfur degassing" and a viable mechanism for the understanding of S exchange between Earth's surface and the lower mantle for the deep sulfur cycle.

## INTRODUCTION

Sulfur (S) is one of the major multi-valent volatile elements, broadly distributed throughout the Earth, and participating in a variety of fundamental geochemical processes (e.g., the global biochemical circulation [1], metal transport [2], atmospheric S loading during the volcanic eruption, and core–mantle segregation [3]). Generally, the chemical speciation of S is strongly influenced by a wide range of oxidation states available. Under highly reducing environments, S dominantly exhibits an oxidation state of -II as sulfide, whereas under strongly oxidizing conditions it shows an oxidation state of +VI in sulfate. Other chemical species where S takes up intermediate oxidation states such as polysulfides, elemental S, sulfite, or thiosulfate sulfite may exist as well in different geochemical settings [2,4,5]. It happens often that the behavior of S in natural processes associated with complex oxidation-reduction reactions is unpredictable due to changes in the oxidation state of S across the range -II to +VI. Therefore, the geochemical behavior of S in the Earth is replete with paradoxes, and there are many open questions in geochemical processes related to S-bearing minerals.

A well-known geological paradox called "sulfur excess degassing" has been evidenced at numerous subduction zone volcanoes [6,7], where the amount of S (principally in the form of $SO_2$) released during explosive eruptions can be orders of magnitude larger than that estimated to degassing from the erupted melt [5]. A variety of sources for the excess S released by magmas in volcanic emission [8,9] have been proposed: dissolution in the silicate liquid [10,11] or a coexisting gas phase at depth before eruption [7,12], gas expulsion from magma mixing [13,14], crystallization-induced exsolution (second boiling) [15], or the breakdown of S-bearing minerals [16]. These mechanisms were proposed based on the magmatic systems, which are related to volcanic eruptions in a shallow crust. It should be noted, however, that the ultimate source of the S found near the Earth's surface is derived from the Earth's mantle [17]. Oxygen is one of the most abundant elements in Earth and has provided critical control on the nature of Earth S reservoirs. The compounds formed by S and O have important implications for the geochemical processes and the nature of Earth S reservoirs. Thus, a key question that needs to be resolved regards the formation and properties of S-O compounds under mantle conditions.

Various S-O compounds such as SO [18], $SO_2$ [19], $SO_3$ [20], $S_7O$ [21], and $S_8O$ [22] have been proposed at ambient pressure. High pressure as a characteristic for the mantle can drastically modify chemical properties of elements and promote the formation of unexpected minerals [23–26]. Currently, several high-pressure $SO_3$ phase have been proposed in theory [27], and only $SO_2$ has been experimentally studied up to 60 GPa [28]. The other S-O compounds have been insufficiently understood at high pressures till now. A pressing task is to investigate the

S-O compounds viable under pressure conditions relevant to Earth's mantle.

Here, we report an extensive exploration of the high-pressure phase diagrams of S-O compounds. Besides the known $SO_2$ and $SO_3$ compounds, an unexpected stoichiometry of $S_3O_4$ with an intriguing crystal structure, which contains a mixture of +II and +IV oxidation states of S, is predicted to appear at high pressure. We show that $S_3O_4$ is able to be produced in reactions of sulfates and sulfides with iron and goethite under high pressure conditions in the deep mantle, and decomposes into $SO_2$ and S at low P-T conditions relevant to shallow depths of Earth, thus offering insightful implications for S cycles, and the origin of excess S degassing observed in volcanic eruptions.

## METHODS

The crystal structure searches on $S_xO_y$ ($x$=1-3, $y$=1-4) at the selected pressures of 50, 70, and 100 GPa have been performed using the swarm intelligence based-CALYPSO method [29–31], which has been successful in resolving crystal structures of a large number of materials at high pressure [32]. The maximum simulation cell of structure searches contains 40 atoms for each composition. Structural optimization, electronic structure, and phonon calculations were performed in the framework of density functional theory within the generalized gradient approximation [33] as implemented in the VASP code [34]. The electron-ion interaction was described by the projector augmented-wave potentials [35], with $3s^23p^4$ and $2s^22p^4$ configurations treated as the valence electrons of S and O, respectively. A kinetic cutoff energy of 900 eV and a spacing of $2\pi \times 0.03$ Å$^{-1}$ for Monkhorst-Pack k-mesh sampling [36] were adopted to give well converged total energies (~1 meV/atom). The ionic positions were fully relaxed until the residual force acting on each ion was less than 1 meV/Å. Due to the layered structure of $S_3O_4$, influence of van der Waals interactions was considered using the optB88-vdW functional [37]. The dynamic stability of the predicted new phases was verified by phonon calculations using the direct supercell method as implemented in the PHONOPY code [38].

## RESULTS

Our main structure searching results are depicted in the convex hull diagrams of Fig. 1(a). The energetic stabilities of a variety of S-O structures are evaluated from their formation enthalpies relative to the dissociation products of the relevant elemental S [39,40] and O solids [41]. At 50 and 70 GPa, the known stoichiometries $SO_2$ and $SO_3$ are readily identified as stable in our structure searching simulations. At 100 GPa, an unexpected composition of $S_3O_4$ becomes stable with respect to the dissociation products of the elemental S and $SO_3$. The predicted stable pressure ranges for the considered structures are listed in Fig. 1(b). $SO_3$ is found to be the most stable phase against decomposition throughout the studied pressure range (50-100 GPa). $S_3O_4$ is energetically favorable relative to decomposition into element S and $SO_2$ or $SO_3$ in the pressure range of 79-102 GPa (Fig. 1c). The emergence of $S_3O_4$ leads to the instability of $SO_2$ above 81.5 GPa. We calculated phonon dispersions and observed no imaginary frequencies for the $S_3O_4$ structures (Fig. S1), indicating that these predicted structures are dynamically stable.

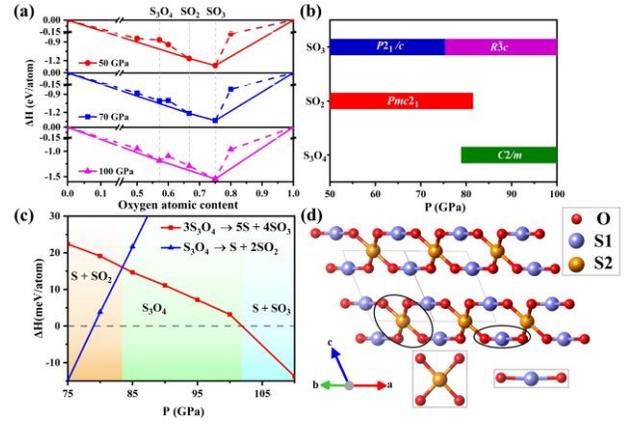

Figure 1: Relative thermodynamic stability of S-O system at 0 K. (a) Convex hull data of the $S_{1-x}O_x$ system at 50 GPa (top), 70 GPa (middle) and 100 GPa (bottom). The formation enthalpies ΔH for each structure were calculated with respect to elemental S and O solids by $\Delta H(S_{1-x}O_x) = H(S_{1-x}O_x) - (1-x)H(S\ solid) - xH(O\ solid) (0 < x < 1)$. The known S-III [39], S-IV [40] and ε-$O_2$ phase [41] selected as the reference structures in the corresponding stable pressure ranges. The stable structures locate on the solid lines, while the metastable structures sit on the dashed lines. (b) Predicted pressure-composition phase diagram of S-O phases. (c) Calculated pressure-enthalpy diagram for the reactions $3S_3O_4 \rightarrow 4SO_3+5S$ and $S_3O_4 \rightarrow 2SO_2+S$ using optB88-vdw functional. The zero-point energy was included in the above energy calculations. (d) Crystal structure of $C2/m$-$S_3O_4$ contains mixed four-fold (S1) and six-fold (S2) coordination of S.

The structure of $S_3O_4$ (Fig. 1d) is inherently layered and contains mixed four-fold and six-fold coordination of S. Specifically, S1 is linearly coordinated to two O atoms, while S2 is square-coordinated to four O atoms. All S atoms are bonded to two adjacent S atoms, thus forming zigzag polymeric all-S chains. The S1-S1 and S1-S2 bond lengths are 2.22 and 2.13 Å at 80 GPa, respectively, slightly longer than the S-S bond lengths (2.01 Å) in the S-III phase, therefore indicating relatively weaker covalent S–S bonding. To further decipher the nature of the bonding, we have examined the electron localization function (ELF) [42] of $S_3O_4$ in the (100) and (010) planes (Fig. 2a). Two inequivalent S atoms are clearly seen, while a less localized charge distribution is seen on the S–O bonds, indicating a significant degree of ionicity between the O anions and S cations. Clear covalent S-S bonding is evidenced by the strong charge localization between the nearest-neighbor S-S.

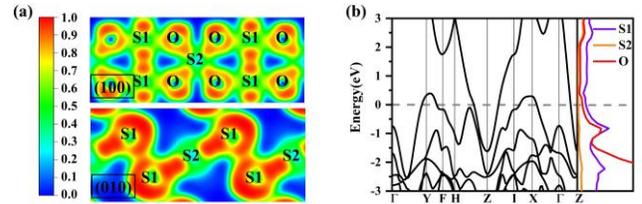

Figure 2. (a) Calculated ELF in (100) and (010) planes of $C2/m$-$S_3O_4$ at 80 GPa. (b) Band structures and the projected density of states (PDOS) of $C2/m$-$S_3O_4$ at 80 GPa. The dashed line indicates the Fermi level. The band structures are calculated using the Heyd–Scuseria–Ernzerhof hybrid functional [43,44].

The oxidation states of S in geological environments play pivotal roles in deciding planetary chemical and physical dynamics [45]. Generally, the oxidation state of element is closely related to the local coordination and charge transfer. The S oxidation states in $SO_2$ and $SO_3$ can be assigned unambiguously as +IV and +VI, respectively. In contrast, the two- and four-fold coordination of S atoms with O atoms in $S_3O_4$ reveals its mixed-valence state. A bader charge analysis [46], summarized in Table 1, corroborates this interpretation. Note that the bader charges systematically underestimate the formal charge state ($O^{2-}$ here has a charge $-1.28e$ in the $SO_2$). In $SO_3$, S has a formal charge state of +VI, and there is a charge transfer of 3.90 $e$ from S to O, very close to that in $SF_6$ (~3.73 $e$). In $SO_2$, S has a formal charge state of +IV, the charge transfer is 2.56 $e$. In $S_3O_4$, the partial charge of 2.68 $e$ in square coordinated S2 is almost equal to that in the $SO_2$ case, so that S2 can be considered as having an oxidation state of +IV. On the other hand, S1 is significantly less positively charged (1.04 $e$) than the $S^{+4}$ anion in $SO_2$. This result highlights a crucial distinction of the S1 compared to S in $SO_2$, indicating that the linearly coordinated S1 in $S_3O_4$ adopts the rare +II S oxidation state.

Table 1. Partial charges for various S-O compounds obtained from bader integration at 80 GPa.

| Compounds | S ($e$) | | O ($e$) |
| --- | --- | --- | --- |
| | S1 | S2 | |
| $S_3O_4$ | +1.04 | +2.68 | -1.19 |
| $SO_2$ | | +2.56 | -1.28 |
| $SO_3$ | | +3.90 | -1.30 |

The S-O compounds tend to be insulating, as satisfying the octet rule usually leads to the opening of a band gap. This rule is applicable to the predicted polymeric phases of $SO_2$ and $SO_3$. However, in $S_3O_4$ two bands are found to cross the Fermi level, forming an electron pocket around the Z point and a hole pocket spanning the X and Y points (Fig. 2b), giving rise to a clear metallic character of $S_3O_4$. The projected density of states (Fig. 2b) shows that both O and the linearly coordinated S1 contribute to the density of electronic states at the Fermi level, and the latter contribution is dominant. The metallic character originates from an overlap of the S1 electron lone pairs, which depends on interlayer distance (Fig. S2).

Since both the S and O are typical light elements, the stability of S-O compounds may be quite sensitive to temperature. To assess the viability at high temperature, we further examine their energetic and structural stability at relevant simultaneous high P-T conditions. The free energies including the vibrational contributions and entropic effects are evaluated for each phase using the quasi-harmonic approximation. Formation enthalpy calculations further reveal that $S_3O_4$ is energetically favorable relative to decomposition into $SO_3$ and S above 70 GPa, and temperature has a minor impact on the threshold pressure (Fig. 3a). Against decomposition into $SO_2$ and S, the stability region of $S_3O_4$ is shifted to higher pressure with rising temperature, increasing from 79 GPa at 0 K to 100 GPa at 2,300 K (Fig. 3b). *Ab-initio* molecular dynamics calculations show that $S_3O_4$ remains firmly solid at 2,000 K in the pressure range 80-100 GPa corresponding to deep mantle conditions (see Fig. S3), revealing that $S_3O_4$ may exist in solid form in the deep mantle. Overall, the predicted $S_3O_4$ is stable at P-T conditions relevant to Earth's lower mantle [47], but decomposes into $SO_2$ and S at low pressure.

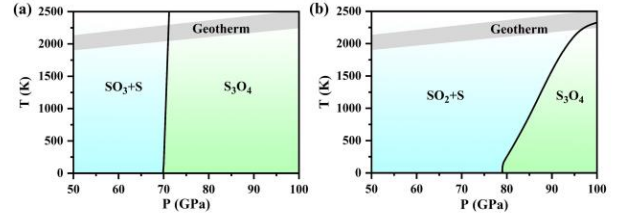

Figure 3. The P-T phase diagrams for the reactions of (a) $4SO_3+5S \rightarrow 3S_3O_4$ and (b) $2SO_2+S \rightarrow S_3O_4$. The geotherm curve is adapted from Ref. [47]

It is well-known that the exchange of S between Earth's surface and the mantle, i.e., transporting S to the mantle via subduction and returning it to the surface by volcanic degassing, results in a global S cycle [48]. The global S cycle involves the transformations of S species via redox-driven chemical processes (e.g., sulfate reduction and sulfide oxidation). Previous studies have indicated that sulfates (e.g., $CaSO_4$ and $MgSO_4$) [48,49] and sulfides (e.g., FeS and $FeS_2$) [50,51] may exist in Earth's mantle. It is estimated that ~1 weight % metallic Fe is present due to self-reduction reactions in lower mantle [52]. Also, the pyrite-type FeOOH could still survive at the P-T conditions of the lower mantle within the deeply subducted slabs [53]. Thus, we explore the possibility to produce $S_3O_4$ by decomposition of sulfates using Fe as a reducing agent or oxidation reactions of sulfides with the FeOOH as an oxidizing agent. We have explored 39 possible reaction routes, in which the iron oxides and hydrogen-bearing minerals that possibly exist in the interior of Earth were chosen as products. The calculated negative reaction enthalpies of seven routes support the formation of $S_3O_4$ via redox reactions at 100 GPa (Fig. 4a), which is relevant to the deep mantle pressure.

According to our results, three possible processes can occur for the S cycle in the Earth (Fig. 4b). Firstly, the S-carrying sulfates or sulfides (e.g., $CaSO_4$, $FeS_2$ [54], and FeS [55]) are transported to the deep mantle within subduction slabs. Then they can react with Fe or FeOOH (minerals present in Earth's mantle) to produce $S_3O_4$ at reducing or oxidizing conditions. If $S_3O_4$ formed in the deep mantle ascends, by mantle dynamic processes, to shallow depths of Earth and low pressure conditions, it would decompose into S and $SO_2$, which is the principal form of S released during explosive eruptions (Fig. 3c). While the mechanism of direct decomposition of S-bearing minerals (e.g., $CaSO_4$, $FeS_2$, and FeS) to release $SO_2$ as an explanation of the "sulfur excess" paradox is not supported even at high pressure [11] (see Fig. S4 for enthalpy calculations), the present compound $S_3O_4$, which was not previously considered, provides an alternative S reservoir in the deep mantle, completes the deep S cycle and helps to explain the paradox of "sulfur excess" in volcanic eruptions.

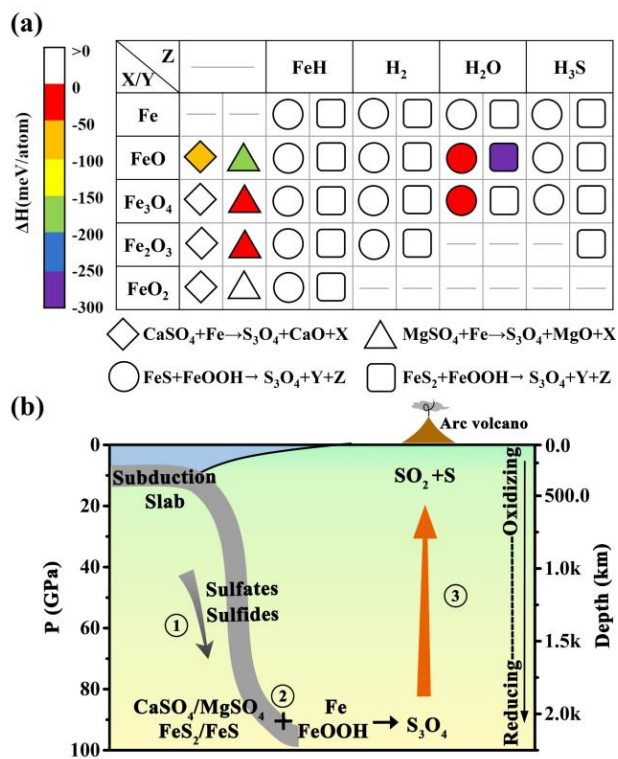

Figure 4. (a) Relative enthalpy of proposed reactions forming $S_3O_4$ at 100 GPa. The rhombus and triangle represent the reactions of $CaSO_4$ and $MgSO_4$ with iron, respectively. The circle and square represent the reactions of FeS and $FeS_2$ with FeOOH, respectively. The crystal structures adopted to evaluate enthalpies are presented in Table S2. (b) The proposed processes for exchange of S between Earth's surface and mantle.

## CONCLUSION

In summary, a hitherto unknown compound $S_3O_4$ has been identified to become stable at high P-T conditions relevant to the deep mantle. It contains a mixture of S(II) and S(IV) oxidation states and exhibits a peculiar metallic nature. A systematic examination of relevant formation and decomposition reactions reveals that $S_3O_4$ might be considered as a key ingredient to promote redox reactions of sulfate or sulfide in the deep mantle and to release $SO_2$ at shallow depths of Earth, thereby offering insightful implication on the origin of excess S degassing observed in volcanic eruptions. The present results have fundamental significance and implications for practical processes in chemistry and geoscience, and further experimental exploration is highly expected.

## ASSOCIATED CONTENT

The supporting information is available free of charge.

Phonon dispersion relations, band gap analysis, mean-square-deviation, calculated enthalpies and Gibbs free energy with different exchange-correlation functionals at high P-T, and pseudopotential measurement of the $S_3O_4$; crystallographic information of S-O compounds and other reactants.

## AUTHOR INFORMATION

†These authors contributed equally to this work.

**Corresponding Author**

*lvjian@jlu.edu.cn;

*maohk@hpstar.ac.cn;

*wyc@calypso.cn.

**Notes**

The authors declare no competing financial interest.

## ACKNOWLEDGMENT

This research was supported by the National Key Research and Development Program of China under Grant No. 2016YFB0201201; the National Natural Science Foundation of China under Grants No. 11404128, 11822404, 11534003 and 21873017; Program for JLU Science and Technology Innovative Research Team; and the Science Challenge Project, No. TZ2016001. Part of the calculation was performed in the high-performance computing center of Jilin University.

Supporting Information for

# Stabilization of $S_3O_4$ at High Pressure-Implications for the Sulfur Excess Paradox

Siyu Liu[1, †], Pengyue Gao[1, †], Andreas Hermann[2], Guochun Yang[3], Jian Lv[1,*], Yanming Ma[1, 4], Ho-Kwang Mao[5, 6, *], and Yanchao Wang[1, *]

[1] State key laboratory of superhard materials & International center of computational method and software, College of Physics, Jilin University, Changchun 130012, China

[2] Centre for Science at Extreme Conditions and SUPA, School of Physics and Astronomy, The University of Edinburgh, Edinburgh EH9 3FD, United Kingdom

[3] State Key Laboratory of Metastable Materials Science & Technology and Key Laboratory for Microstructural Material Physics of Hebei Province, School of Science, Yanshan University, Qinhuangdao 066004, China

[4] International Center of Future Science, Jilin University, Changchun 130012, China

[5] Center for High Pressure Science and Technology Advanced Research, Beijing 100094, China.

[6] Geophysical Laboratory, Carnegie Institution of Washington, Washington, DC 20015, USA.

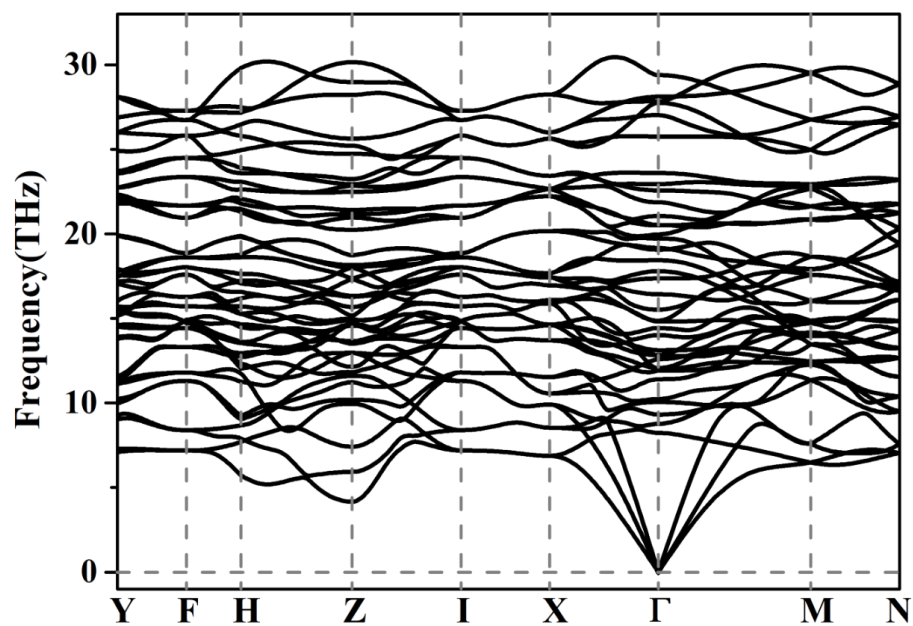

**Figure S1**. Phonon dispersion relations along select high-symmetry points in the Brillouin zone for *C*2/*m*-S$_3$O$_4$ at 100 GPa.

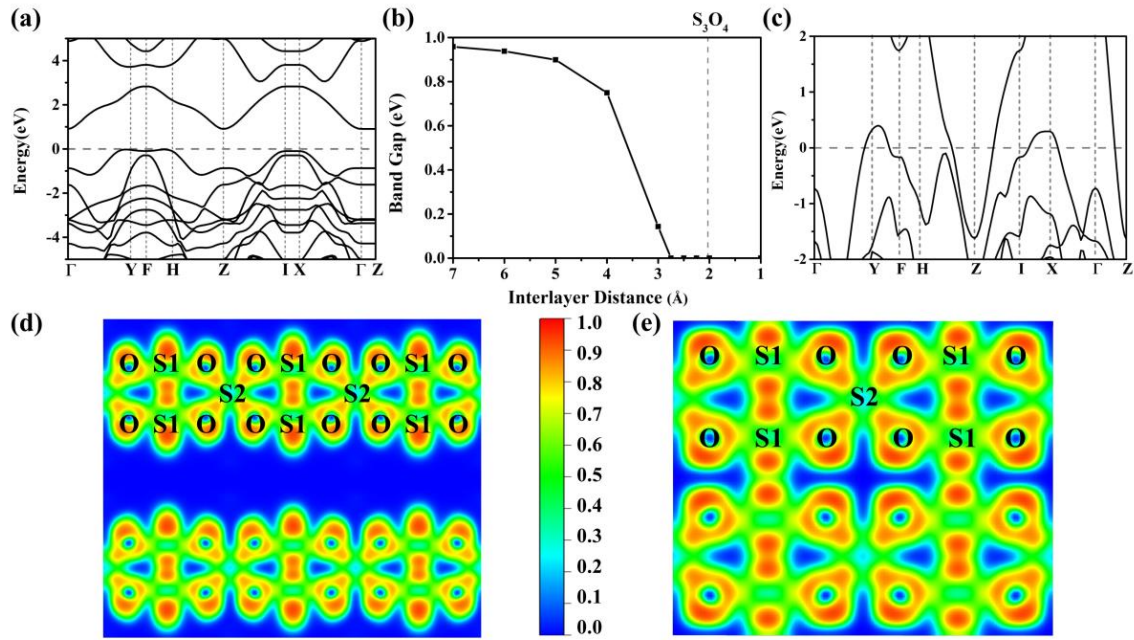

**Figure S2.** (a) Band structure of the $S_3O_4$ with 10 Å interlayer distance. (b) The band gaps as a function of interlayer distance. (c) Band structure of the $S_3O_4$. The horizontal dashed line indicated the Fermi energy level. Calculated electron localization function in (100) plane of (d) $S_3O_4$ with 5 Å interlayer distance and (e) $S_3O_4$ phase. The band gap strongly depends on interlayer distance. The $S_3O_4$ shows the insulating character when the interlayer distance is larger than 2.75 Å. Interestingly, reducing interlayer distance makes part of electrons in the layer transfer into the interlayer regions, enhancing the delocalized degree of electrons and resulting in the metallization.

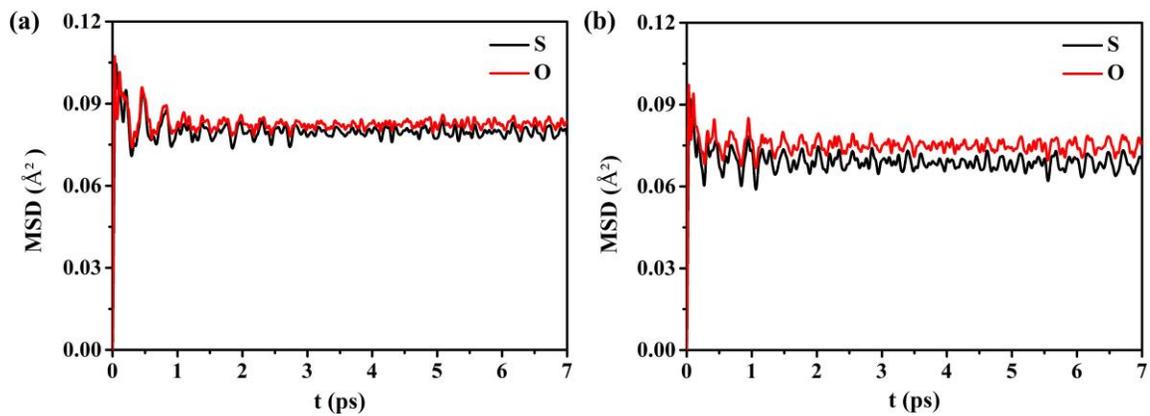

**Figure S3.** Calculated mean-square-deviation of S and O in $S_3O_4$ at 2,000 K at (a) 80 GPa and (b)100 GPa.

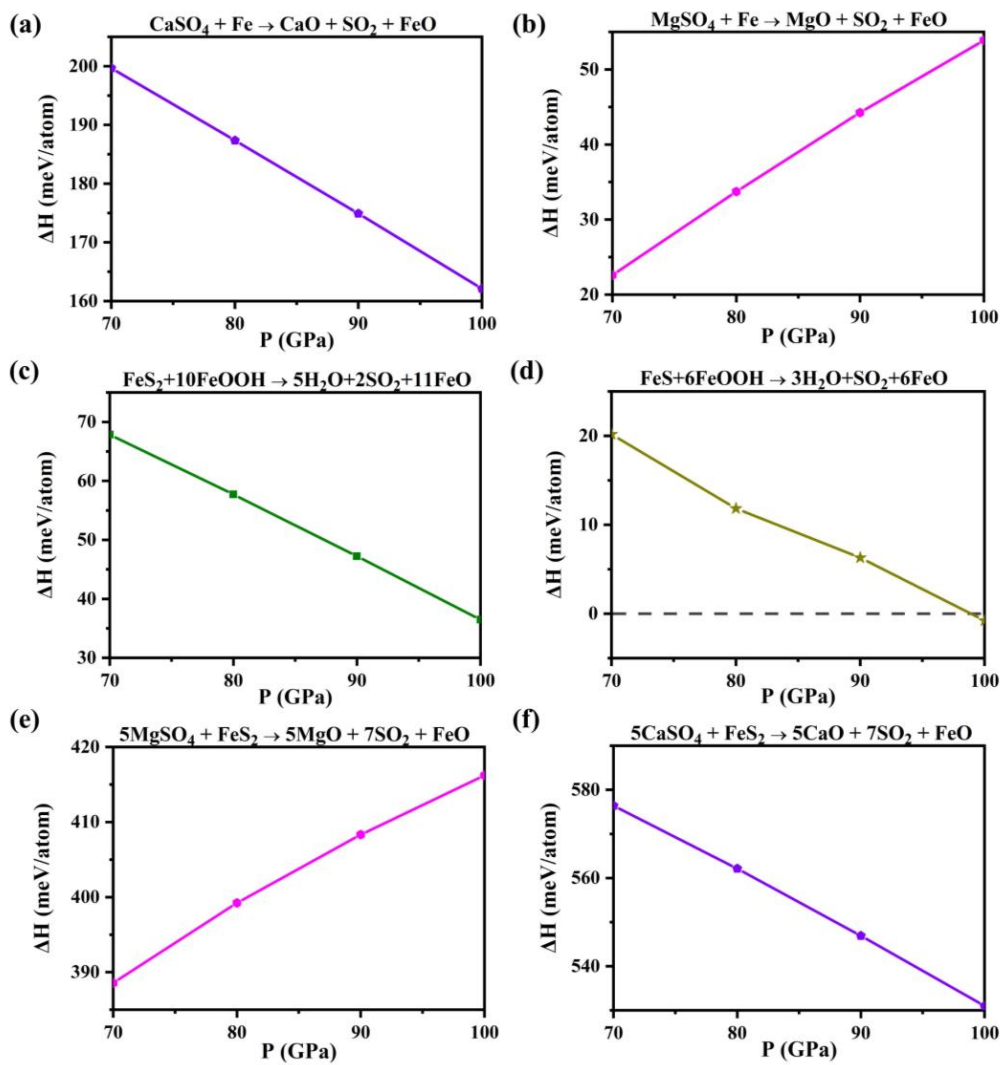

**Figure S4.** Relative enthalpy of some potential reactions forming $SO_2$ at high pressure. (a-d) The direct decomposition of both sulfates ($CaSO_4$/$MgSO_4$) and sulfides ($FeS_2$/FeS) to release $SO_2$ are not supported by the enthalpy calculation. (e-f) The oxidized sulfates and the reduced sulfides cannot directly produce $SO_2$.

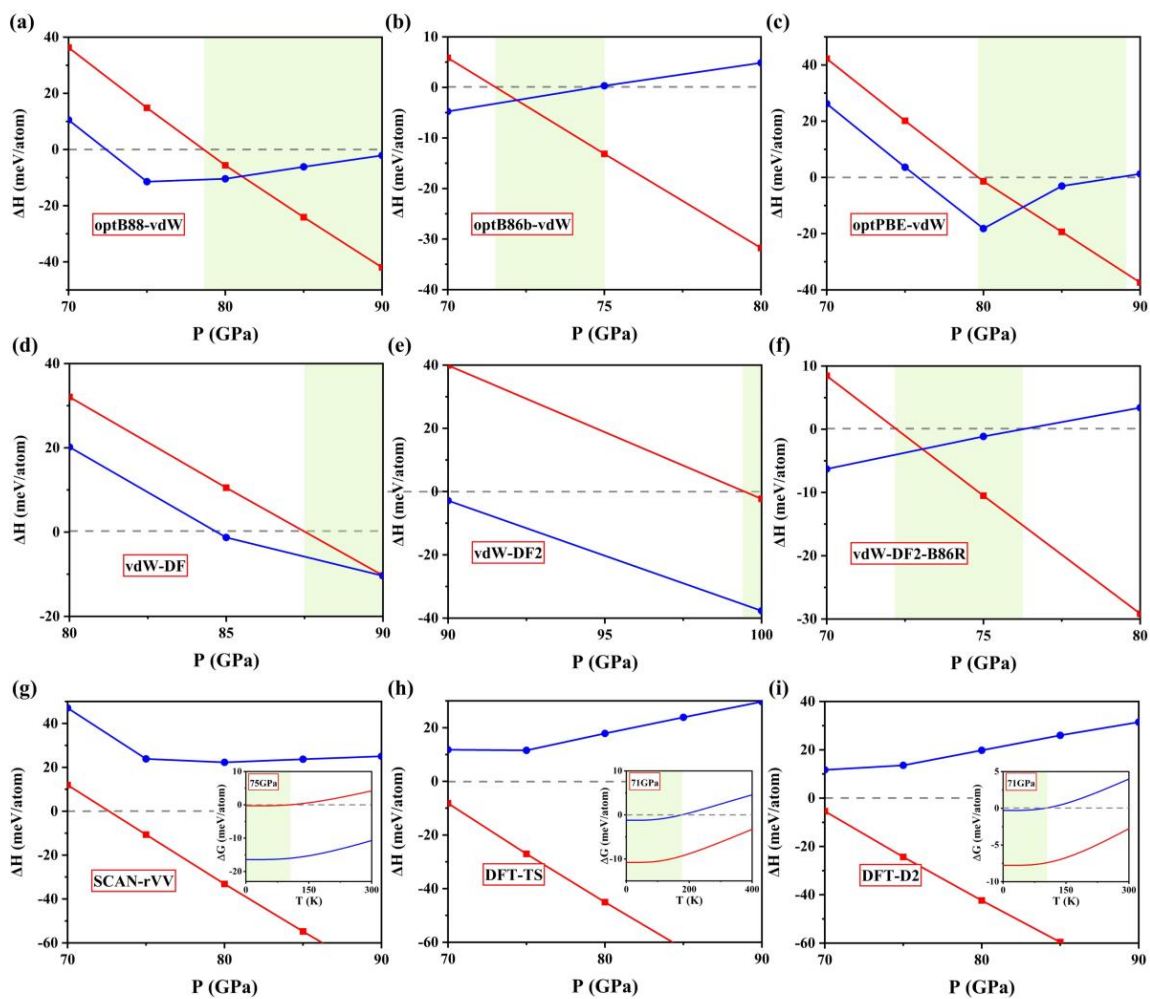

**Figure S5.** Relative enthalpies and Gibbs free energy of proposed reactions for $4SO_3+5S \rightarrow 3S_3O_4$ and $2SO_2+S \rightarrow S_3O_4$ with various van der Waals (vdW) methods at high pressure and temperature (P-T). Both negative energy values of the two reactions reveal that $S_3O_4$ is stable. Due to the typical layer structure of $S_3O_4$, the vdW interactions were considered to inclusion of the dispersion interactions. All the calculated vdW methods give the conclusion that $S_3O_4$ is stable at high P-T.

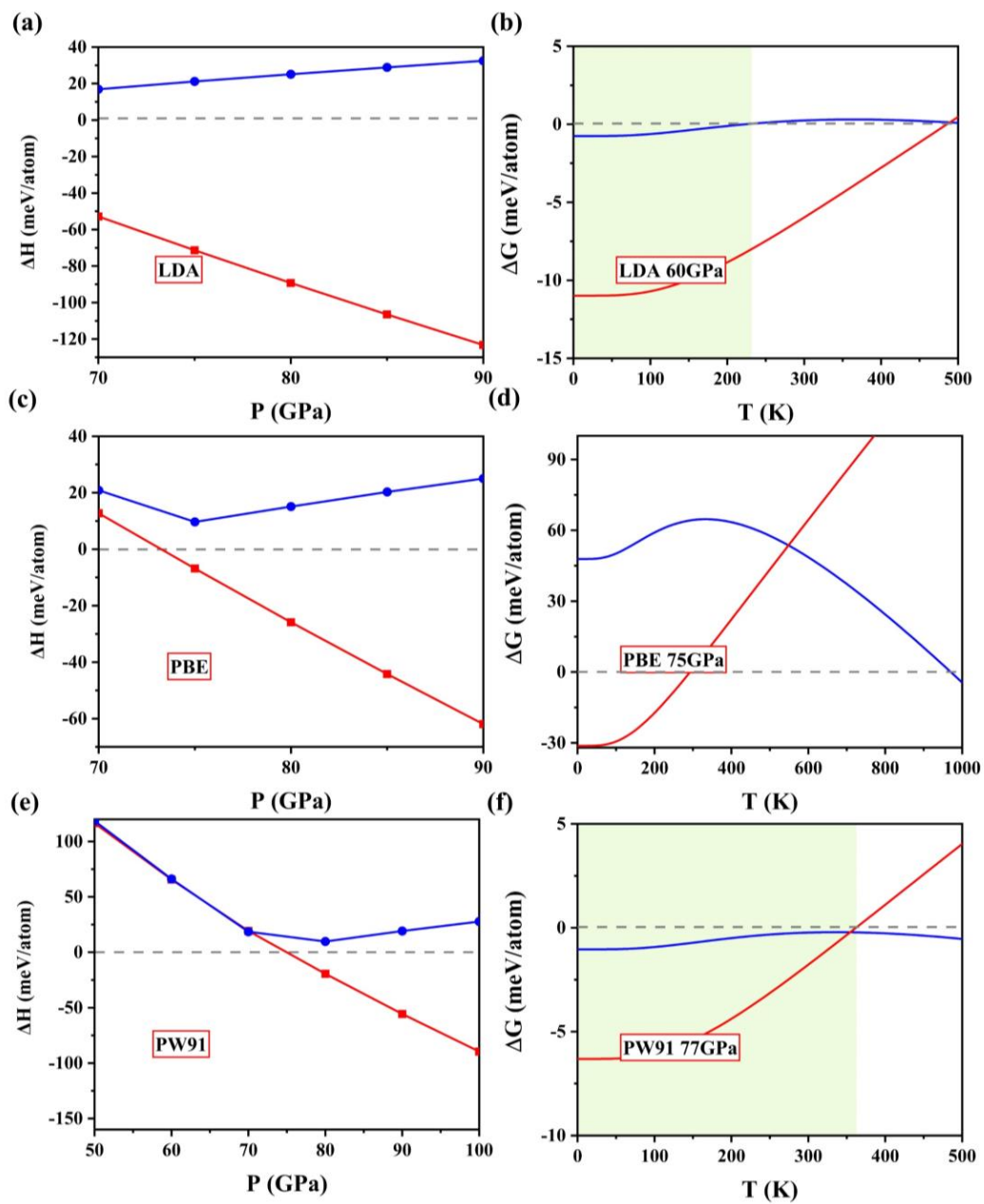

**Figure S6.** The calculated Gibbs free energy of proposed reactions for $4SO_3+5S \rightarrow 3S_3O_4$ and $2SO_2+S \rightarrow S_3O_4$ with three different exchange-correlation functionals at high P-T. Even without the consideration of vdW, the calculations within local density approximation and Perdew-Wang-91 indicate that the $S_3O_4$ is stable at high temperature.

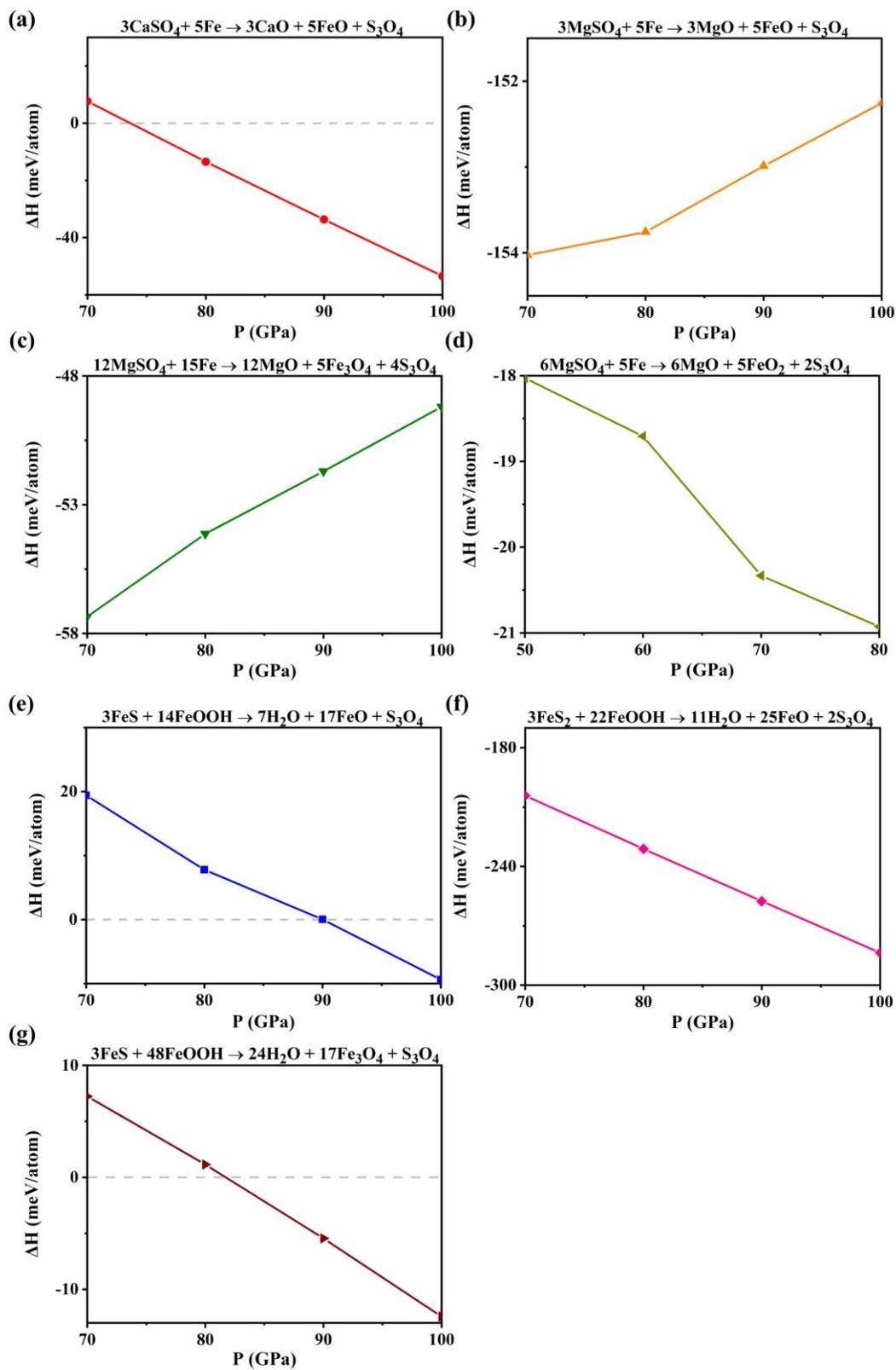

**Figure S7.** Relative enthalpy of proposed reactions forming $S_3O_4$ at high pressure.

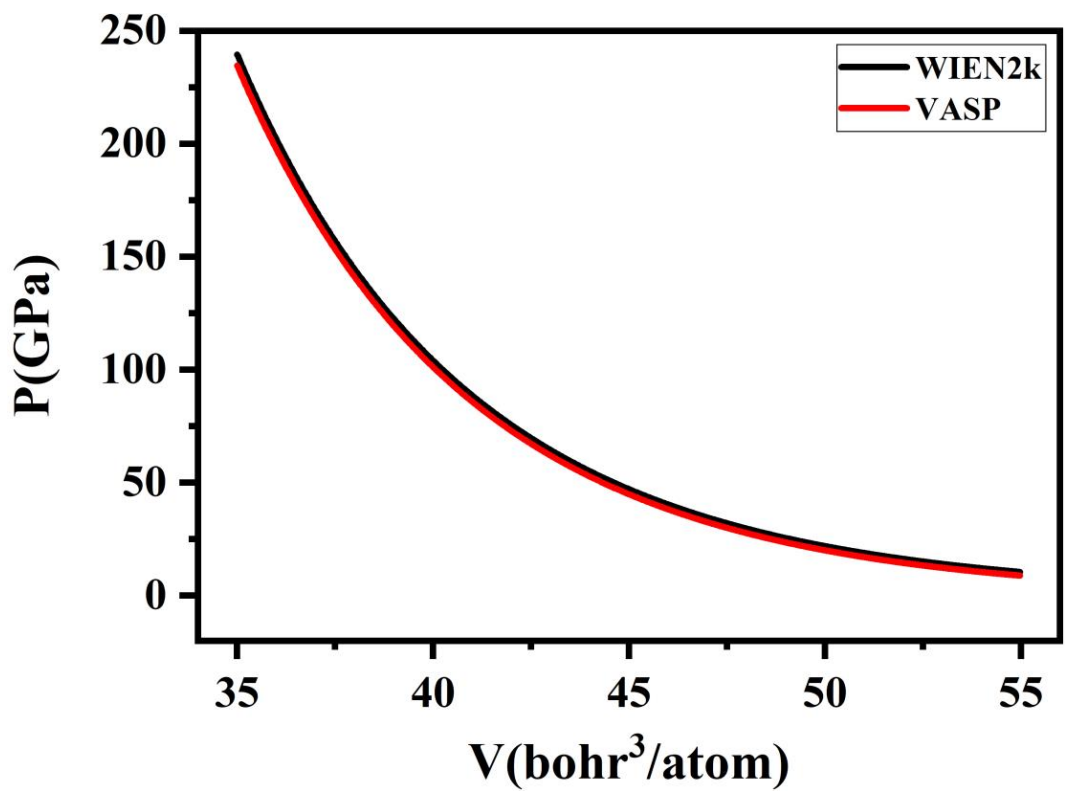

**Figure S8**. Volumes as a function of pressure for $R\bar{3}c$-$SO_3$ calculated by projector augmented-wave (PAW) of VASP and all-electron potential of WIEN2k. The relative error is nearly 2%, validating our calculations within the PAW.

**Table S1.** Detailed crystallographic information for the energetically stable structures of S-O compounds identified in our structure search.

| System | P(GPa) | Space group | Lattice constants (Å,°) | Atom | Atomic coordinates (Fractional) | | |
|---|---|---|---|---|---|---|---|
| | | | | | X | Y | Z |
| $SO_2$ | 50 | $Pmc2_1$ | a=4.4794<br>b=3.9383<br>c=5.6596 | S(2a) | 0.00000 | 0.74077 | -0.56433 |
| | | | | S(2b) | 0.50000 | 0.13662 | -0.45431 |
| | | | | O(2a) | 0.00000 | 0.62093 | -0.01845 |
| | | | | O(2b) | 0.50000 | 0.34177 | -0.24091 |
| | | | | O(4c) | 0.74985 | 0.13991 | -0.87690 |
| $SO_3$ | 100 | $R\bar{3}c$ | a=3.9610<br>c=10.7980 | S(6b) | 0.33333 | 0.66667 | -0.33333 |
| | | | | O(18e) | 0.34160 | 0.00000 | -0.25000 |
| $S_3O_4$ | 100 | $C2/m$ | a=4.6818<br>b=5.5553<br>c=4.6304<br>β=119.1969 | S(4i) | 0.97725 | 0.00000 | 0.74981 |
| | | | | S(2b) | 0.00000 | 0.50000 | 0.00000 |
| | | | | O(8j) | 0.99654 | 0.69447 | 0.24908 |

**Table S2.** The formations of reactants and the products of calculations at 100 GPa.

| System | Space Group | Lattice Constants (Å, °) | Atom | X | Y | Z | Reference |
|---|---|---|---|---|---|---|---|
| CaSO$_4$ | $I4_1/a$ | a=4.2958<br>c=9.9909 | Ca(4b)<br>S(4a)<br>O(16f) | 0.5000<br>0.0000<br>0.2438 | 0.5000<br>0.0000<br>0.1259 | 0.0000<br>0.0000<br>0.0794 | [1] |
| CaO | $Pm\bar{3}m$ | a=2.5687 | Ca(1a)<br>O(1b) | 0.0000<br>0.5000 | 0.0000<br>0.5000 | 0.0000<br>0.5000 | [2] |
| Fe | $P6_3/mmc$ | a=2.2947<br>c=3.9937 | Fe(2c) | 0.3333 | 0.6667 | 0.2500 | [3] |
| Fe$_3$O$_4$ | $Cmcm$ | a=2.6177<br>b=8.9120<br>c=8.8798 | Fe(4c)<br>Fe(8f)<br>O(4c)<br>O(8f)<br>O(4a) | 0.5000<br>0.5000<br>0.5000<br>0.5000<br>0.0000 | -0.3785<br>-0.1385<br>-0.0306<br>-0.1990<br>0.0000 | 0.7500<br>0.9363<br>0.7500<br>0.3952<br>0.0000 | [4] |
| FeO$_2$ | $Pa\bar{3}$ | a=4.3005 | Fe(4a)<br>O(8c) | 0.5000<br>-0.3545 | 0.0000<br>0.8545 | 0.5000<br>0.1455 | [5] |
| FeS$_2$ | $Pa\bar{3}$ | a=4.8879 | Fe(4a)<br>S(8c) | 0.0000<br>0.3826 | 0.0000<br>0.3826 | 0.0000<br>0.3826 | [6] |
| H$_2$O | $Pn\bar{3}m$ | a=2.6589 | H(4b)<br>O(2a) | 0.2500<br>0.0000 | 0.2500<br>0.0000 | 0.2500<br>0.0000 | [7] |
| FeH | $Fm\bar{3}m$ | a=3.4619 | Fe(4a)<br>H(4b) | 0.0000<br>0.5000 | 0.0000<br>0.5000 | 0.0000<br>0.5000 | [8] |
| MgSO$_4$ | $I4_1/a$ | a=4.1343<br>c=9.4376 | Mg(4a)<br>S(4b)<br>O(16f) | 0.0000<br>0.5000<br>0.7563 | 0.0000<br>0.5000<br>0.6460 | 0.0000<br>0.0000<br>0.0792 | [9] |
| MgO | $Fm\bar{3}m$ | a=3.796 | Mg(4a)<br>O(4b) | 0.0000<br>0.0000 | 0.0000<br>0.5000 | 0.0000<br>0.5000 | [10] |
| FeO | $R\bar{3}m$ | a=2.8908<br>c=6.3073 | Fe(3a)<br>O(3b) | 0.0000<br>0.3333 | 0.0000<br>0.6667 | 0.0000<br>0.1667 | [11] |
| Fe$_2$O$_3$ | $Cmcm$ | a=2.6136<br>b=8.6846<br>c=6.4257 | Fe(4a)<br>Fe(4c)<br>O(4c)<br>O(8f) | -0.5000<br>-0.5000<br>-0.5000<br>-0.5000 | -0.5000<br>0.2506<br>-0.4070<br>-0.1467 | 0.5000<br>0.2500<br>0.2500<br>0.4308 | [12] |
| FeS | Pnma | a=4.7961<br>b=3.0110<br>c=4.9234 | Fe(4c)<br>S(4c) | 0.0149<br>0.1987 | 0.2500<br>0.2500 | 0.2005<br>0.5864 | [13] |
| FeOOH | Pa$\bar{3}$ | a=4.5518 | Fe(4b)<br>O(8c)<br>H(4a) | 0.5000<br>0.3551<br>0.5000 | 0.5000<br>0.6449<br>0.0000 | 0.5000<br>0.1449<br>0.5000 | [14] |
| H$_2$ | C2/c | a=5.7291<br>b=3.2922<br>c=4.8403<br>β=141.999 | H(8f)<br>H(8f)<br>H(4e)<br>H(4e) | 0.2493<br>0.1445<br>0.0000<br>0.0000 | 0.0984<br>0.1980<br>0.1854<br>0.4045 | 0.7574<br>0.7691<br>0.2500<br>0.2500 | [15] |
| H$_3$S | C2/c | a=7.915<br>b=4.5907<br>c=14.2041<br>β=146.3452 | H(8f)<br>H(8f)<br>H(8f)<br>H(8f)<br>H(8f)<br>H(8f)<br>S(8f)<br>S(8f) | 0.8151<br>-0.5644<br>0.4672<br>-0.0189<br>0.2864<br>0.0609<br>-0.2600<br>0.3294 | 0.3992<br>0.8638<br>0.7567<br>0.3356<br>0.3856<br>0.1667<br>0.6037<br>0.3993 | 0.4903<br>-0.4879<br>0.4193<br>-0.3275<br>0.3267<br>0.3388<br>-0.4222<br>0.2452 | [16] |